\documentclass[%
 reprint,
superscriptaddress,
 amsmath,amssymb,
 aps,
 prl,
floatfix,
]{revtex4-2}

\usepackage{graphicx}
\usepackage{dcolumn}
\usepackage{bm}
\usepackage{amsmath,amssymb}
\usepackage{hyperref}

\hypersetup{
	colorlinks=true,
	linkcolor=blue,
	filecolor=blue,
	urlcolor=blue,
	citecolor=blue,
}

\begin{document}


\title{Achieving the Quantum Limits with Twin-Field Sensors}

\author{Yaohua Li}
\affiliation{%
 Niels Bohr Institute, University of Copenhagen, Jagtvej 155A, Copenhagen DK-2200, Denmark
}%

\author{Fan Yang}
\email{fan.yang@zju.edu.cn}
\affiliation{School of Physics and Zhejiang Key Laboratory of Micro-nano Quantum Chips \\ and Quantum Control, Zhejiang University, Hangzhou 310027, China}
\affiliation{Niels Bohr International Academy, Niels Bohr Institute, University of Copenhagen, DK-2100 Copenhagen, Denmark}

\author{Klaus Mølmer}%
 \email{klaus.molmer@nbi.ku.dk}
\affiliation{%
 Niels Bohr Institute, University of Copenhagen, Jagtvej 155A, Copenhagen DK-2200, Denmark
}%

\date{\today}

\begin{abstract}
Continuous measurements play a fundamental role in quantum experiments, yet understanding and extracting the maximal possible information in a continuous probe signal remains a significant challenge. In this work, we show that for a wide class of sensors, the quantum Fisher information bound can be saturated through a combined measurement of signals from the original sensor and a suitable system copy. We elucidate the physical mechanism of such a twin-field sensor (TFS) and study its performance across diverse implementations, ranging from two-level atoms to a hybrid quantum Rabi model. In all of these cases, we find that simple photon counting or homodyne detection saturates the theoretical quantum limit. We further prove that the TFS achieves the quantum limits for the joint sensing of pairs of parameters as quantified by the quantum Fisher information matrix.
\end{abstract}

\maketitle

{\it Introduction}---Improving measurement sensitivity is a central goal of quantum science and technology \cite{giovannetti_quantum_2006,schnabel_squeezed_2017,braun_quantum-enhanced_2018,pezze_quantum_2018,li_cavity_2021,huang_entanglement-enhanced_2024,demille_quantum_2024}. Driven dissipative quantum systems leak information about their dynamics through radiation continuously emitted into the environment \cite{albarelli_ultimate_2017,shankar_continuous_2019,rossi_noisy_2020,ilias_criticality-enhanced_2022,fallani_learning_2022,nurdin_parameter_2022,patra_single-shot_2022,cabot_continuous_2024,yang_quantum_2026}. By monitoring this output field over time, one obtains a record of measurement data which are subject to the randomness of quantum measurements and are correlated due to the effect of measurement back action on the emitter system.

When extracting information from a continuous measurement, Bayes' rule yields the optimal estimation of unknown system parameters from a given measurement record \cite{gambetta_state_2001,gammelmark_bayesian_2013}, while the Classical Fisher Information (CFI) quantifies the average precision of the measurement over many realizations. The Quantum Fisher Information (QFI) represents the quantum limit of extractable information through the dependence of the full quantum state on the unknown parameters and yields the maximum value of the CFI over all measurement strategies  \cite{gammelmark_fisher_2014}. Achieving this maximum in a real measurement is a highly nontrivial task, especially when it involves probing of the field emitted by a sensor system \cite{godley_adaptive_2023}, as one must identify optimal measurements on the high-dimensional combined Hilbert space of the emitter and the quantized field. As a result, optimal measurements may need to address intricate entanglement within the field and emitter, making a practical implementation difficult.

Despite the general complexity of continuous sensing, there are cases where simple photon counting or homodyne detection can reach the quantum limit. For example, in resonance fluorescence where a two-level atom is driven resonantly with an unknown Rabi frequency $\Omega$, it is shown in Ref.~\cite{gammelmark_fisher_2014} that photon counting achieves high sensitivity due to characteristic oscillations in the waiting time distribution between detector clicks. Homodyne detection with a real local oscillator phase achieves the same sensitivity, but with important contributions from multi-time correlations in the signal record \cite{kiilerich_bayesian_2016}. More recently, it was shown that similar measurement protocols can reach the quantum limit for a class of resonantly driven spin-boson models and that a key requirement for this to happen is a constant phase relation between the state amplitudes of the sensor during the evolution \cite{mattes_designing_2025}.

\begin{figure}[b]
    \centering
    \includegraphics[width=\linewidth]{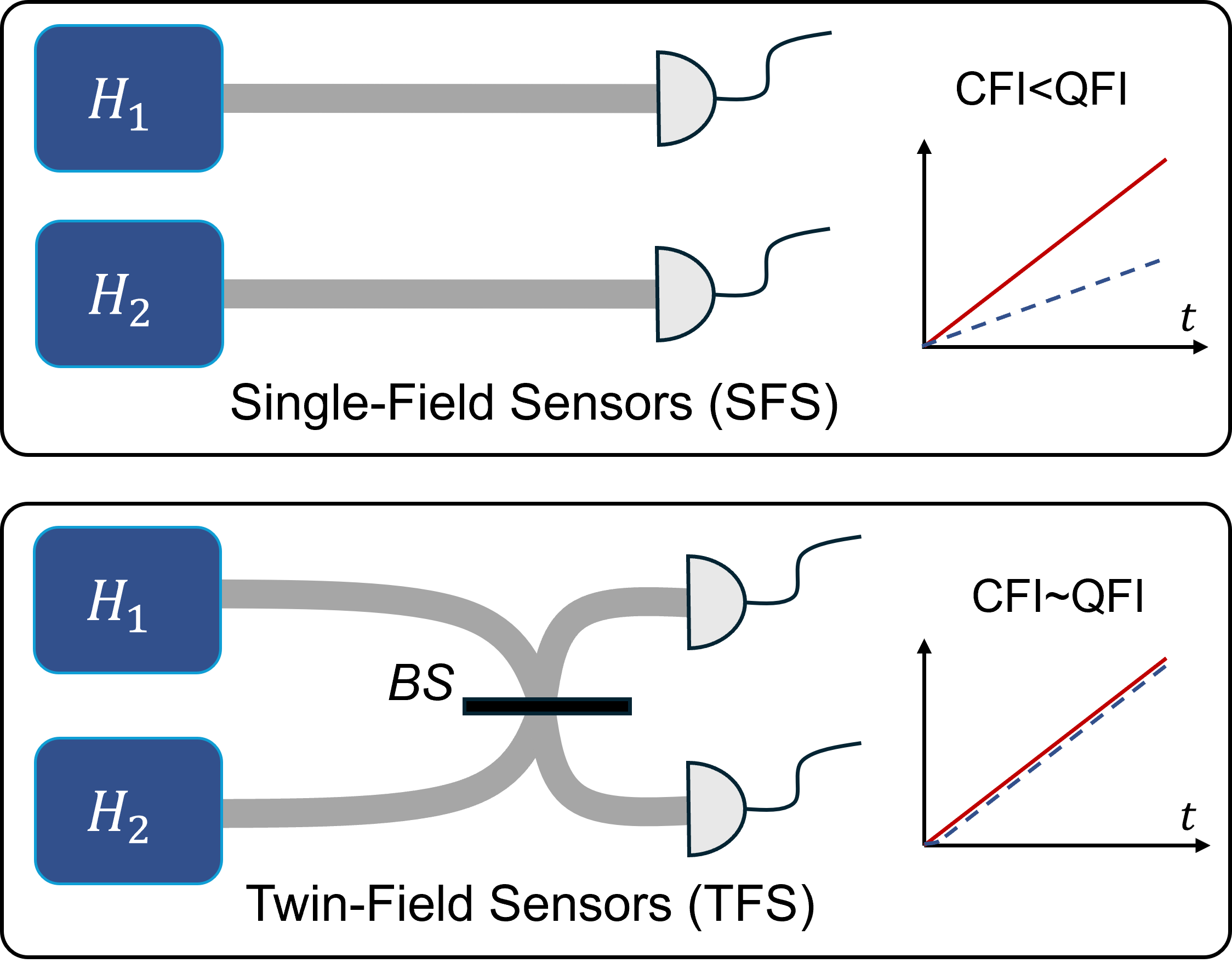}
    \caption{Single-field sensors (SFS, upper panel) directly measure the emitted fields of two systems, while the twin-field sensors (TFS, lower panel) first mix the fields on a beam splitter (BS). We show that if we can implement system Hamiltonians $H_1$ and $H_2$ with opposite detunings and {symmetric encoding} of the parameter of interest, the classical Fisher Information of the TFS detection, quite generally, reaches the quantum Fisher Information, i.e., the theoretical optimum over any detection protocol.}
    \label{fig:scheme}
\end{figure}
In fact, if the sensor system is excited resonantly with a proper phase convention (in a basis $\{|\alpha\rangle\}$), by expanding the emitted field in discrete time-bin Fock space $|\{n_i\}\rangle=|n_1,\cdots,n_N\rangle$, the joint state $|\psi\rangle = \sum_{\alpha,\{n_i\}} c_{\alpha,\{n_i\}}|\alpha\rangle\otimes |\{n_i\}\rangle$ will only have real amplitudes $c_{\alpha,\{n_i\}}$.
However, if the amplitudes $c_{\alpha,\{n_i\}}$ are not restricted to be real during the evolution, the varying phases of the amplitudes $c_{\alpha,\{n_i\}}$ can carry information about the driving field, which hinders the saturation of the quantum sensitivity limit by photon or homodyne detection. This happens, e.g., for off-resonant driving. In this Letter, we establish a continuous sensing architecture that can restore the constant phase relationship under general circumstances and allows classical measurements of the emitted field to saturate the quantum limit.

{\it The twin-field sensor}---Our strategy is to construct a twin-field sensor (TFS). As illustrated in the lower panel of Fig.~\ref{fig:scheme}, supposing the original sensor system is described by a Hamiltonian $H_1$, we introduce an auxiliary twin system subject to a similar, but modified, Hamiltonian $H_2$, such that the joint Hamiltonian can be transformed into a pure-imaginary form, which keeps the combined system-twin-environment state real during the measurement dynamics. For a wide class of Hamiltonians, we find that a simple frequency inversion in $H_2$ and a balanced beam-splitting transformation suffice to achieve the desired condition, making the experimental implementation quite feasible, as one just needs to perform a combined measurement of signals emitted from two independent systems. When sensing parameters are symmetrically encoded in the Hamiltonian operators $H_1$ and $H_2$, we demonstrate that the TFS configuration saturates the QFI sensitivity limit by continuous photon counting or homodyne detection.

To gain an intuitive understanding of how the TFS works and why it can outperform the single-field sensor (SFS, see the upper panel of Fig.~\ref{fig:scheme}), we first consider a simple example: sensing the frequency shift of an optical cavity. Here, directly counting the photons from a leaking cavity prepared in a coherent or Fock state yields no information about the frequency of the cavity field. However, if one can prepare two cavities that are subject to opposite detunings $\pm \delta$ about a common frequency $\omega_0$, mixing the signals on a beam splitter gives rise to a beat note that directly reveals the value of $\delta$.

{\it Theory}---Let us now present a quantitative analysis of the TFS. Supposing that the individual sensor system components ($n=1,2$) are leaking into independent Markovian environments as in the SFS, the combined system is governed by a master equation
\begin{equation}
    \dot{\rho}=-i[H_\mathrm{tot},\rho]+\kappa\sum_{n=1,2}\left[c_{n}\rho c_{n}^{\dagger}-\frac{1}{2} \left(c_{n}^{\dagger}c_{n}\rho+\rho c_{n}^{\dagger}c_{n}\right) \right],
    \label{eq:master}
\end{equation}
where $H_\mathrm{tot}=H_1+H_2$ is the total Hamiltonian, $\kappa$ is the decay rate, and $c_n$ denotes the jump operators of the two systems, which can be, e.g., bosonic annihilation operators or spin lowering operators. Under continuous photon counting, evolution of the conditional state $|{\psi(t)}\rangle$ of the system is described by a quantum trajectory $|\psi(n\delta t)\rangle\propto K_n \cdots K_2 K_1 |\psi(0)\rangle$, in which the instantaneous stochastic evolution is governed by either the no-jump dynamics or quantum jumps associated with detection of a photon in one of the beams, described by $K_j \in \{1-iH_\mathrm{NH}\delta t,c_1,c_2\}$, with the non-Hermitian Hamiltonian $H_\mathrm{NH}=H_\mathrm{tot}-i\kappa(c_1^\dagger c_1+c_2^\dagger c_2)/2$ \cite{dalibard_wave-function_1992}.

The key mechanism of the TFS is to ensure that all Kraus operators are real-valued up to a global phase, such that the conditional states do not acquire a time-dependent phase and, equivalently, the amplitudes of the joint state of the system and the emitted field remain real over time, thus satisfying the criterion of Ref.~\cite{mattes_designing_2025}. 
In the TFS configuration, a beam splitter mixes the fields emitted from two sensors, effectively applying a unitary transformation of the jump operators, e.g., $c_{1,2}\rightarrow c_\pm=(c_1\pm c_2)/\sqrt{2}$. The master equation Eq.~\eqref{eq:master} is invariant under such a transformation, but the quantum trajectory evolution is altered to the one described by
\begin{equation}\label{eq:kraus}
    K_j\in\{1-iH_\mathrm{NH}\delta t,c_+,c_-\}.
\end{equation}

We find that these new Kraus operators can be real-valued for a wide class of Hamiltonians. To illustrate the general mechanism, we consider a single bosonic mode or collective spin, supposing the two systems are described by Hamiltonians $H_{n}(\theta)=f_{n}(\theta,c_{n},c_{n}^{\dagger})$ for $n=1,2$ that satisfy $f_{2}=-f_{1}$, and $\theta$ is the estimated parameter encoded in both systems. Then the total Hamiltonian $ H _ {\mathrm {tot}} $ becomes odd under the exchange of the two systems and thus contains only terms with an odd number of the antisymmetric operators $c_{-}$ $(c_{-}^{\dagger})$. As the two collective operators $c_\pm$ have independent gauge degrees of freedom, we can introduce a unitary transformation such that $Uc_+U^\dagger=c_+$, $U c_-U^\dagger=i\tilde{c}_-$, where $U$ does not depend on $\theta$ and $\tilde{c}_-$ is still a real-valued operator. In this way, all the Kraus operators in Eq.~\eqref{eq:kraus} become real-valued (a global phase factor in front of the jump operator can be neglected). 

As an example, the total Hamiltonian of a detuned cavity $H_{\mathrm{tot}}=-\delta(c_{1}^{\dagger}c_{1}-c_{2}^{\dagger}c_{2})$ together with the non-Hermitian term describing the no-jump dynamics  can be transformed into a pure imaginary form
\begin{equation}\label{eq:fre}
    \tilde{H}_{\mathrm{NH}}    =i\delta(c_{+}^{\dagger}\tilde{c}_{-}-\tilde{c}_{-}^{\dagger}c_{+})-i\kappa(c_+^\dagger c_++\tilde{c}_-^\dagger \tilde{c}_-)/2.
\end{equation}
In fact, for Hamiltonians with complex driving terms $\Omega_{i}^{*} c_{i}^{m}+\Omega_{i}c_{i}^{\dagger m}$, we can introduce a beam-splitting transformation with phase shifts which ensures that we saturate the QFI, as long as $|\Omega_{1}|=|\Omega_{2}|$. A combination of multiple different driving terms, however, will break this freedom. Our strategy can also be extended to hybrid systems with multiple modes and spins, where the same argument applies after introducing suitable unitary transformations for the additional modes or spins.

We note that if photon counting preserves the real-valued state amplitudes, this also holds for homodyne detection of the same mixed fields with an imaginary and a real local oscillator amplitude. Such a homodyne detection leads to a stochastic evolution described by $K_j \propto (1-iH_\mathrm{NH}\delta t) + \sqrt{\kappa}( c_+d\xi_+  +ic_-d\xi_- )$, where $d\xi_{\pm}$ are real-valued back-action amplitudes associated with Wiener noise fluctuations. These operators also maintain the real-valued states after the same unitary transformation (see End Matter for examples).

With the choice of Hamiltonian and the mixed lowering operators associated with the field emitted by the sensor systems, the desired phase relationship between state amplitudes will be automatically attained during the quantum trajectory evolution, but for simplicity, we will assume it is satisfied from the beginning by preparing systems in proper initial states, e.g., the ground (vacuum) states of atomic (optical) systems. Before considering specific examples, we first briefly describe the methods to calculate the QFI and CFI.

{\it Calculation of the QFI---}
We recall that the QFI is directly given by the overlap between pure states in the Hilbert space of the system and the emitted field \cite{braunstein_statistical_1994}, which, for an emitter coupled in a Markovian manner to an environment, can in turn be calculated by solving a generalized master equation for the emitter system of the form \cite{gammelmark_fisher_2014,molmer_hypothesis_2015}
\begin{equation}\label{eq:gm}
    \begin{split}
        \dot{\rho^{\prime}}=&-i\left[H(\theta_{1})\rho^{\prime}-\rho^{\prime}H(\theta_{2})\right]\\
        &+\sum_{n=1,2}\kappa\left[c_{n}\rho^{\prime} c_{n}^{\dagger}-\frac{1}{2} \left(c_{n}^{\dagger}c_{n}\rho^{\prime}+\rho^{\prime} c_{n}^{\dagger}c_{n}\right) \right],
    \end{split}
\end{equation}
where $\theta_1,\ \theta_2$ represent candidate values of the unknown parameter $\theta$ in the Hamiltonian,
and the QFI for the joint system-field state can be obtained as $F_{\theta}=4\partial_{\theta_{1}}\partial_{\theta_{2}}(\log\mathrm{Tr\rho^{\prime}})|_{\theta_{1}=\theta_{2}=\theta}$. 
As Eq.~\eqref{eq:gm} is invariant under the transformation between $c_{1,2}$ and $c_{\pm}$, corresponding to the situations with or without the beam splitter, the value of the QFI equals the one obtained for the two separate emitters. The point of the TFS scheme is not that it offers a quantum sensing scheme with a higher QFI but that it offers a practical protocol that saturates the QFI of the separate systems.

The QFI obtained from Eq.~\eqref{eq:gm} represents the information held within the combined system and emitted fields. As the information in the fields accumulates linearly with time, the QFI of the field (field QFI) is nearly the same as the QFI of the joint system-field state (total QFI) in the long-time limit {for an ergodic sensor} \cite{yang_efficient_2023}, and we aim for the CFI to attain the same asymptotic scaling with time. {For a non-ergodic sensor, the difference between the field QFI and the total QFI can also accumulate with time, so in this case is is relevant to compare the CFI with the field QFI, for which there is also a simple expression $F_{\theta}=-4\partial_{\theta_{2}}^{2}\mathrm{tr}\sqrt{\rho^{\prime}\rho^{\prime\dagger}}|_{\theta_{1}=\theta_{2}=\theta}$ \cite{yang_efficient_2023}.}

{\it Calculation of the CFI---} We now turn to the evaluation of the CFI. For a continuous photon counting measurement, we apply Monte Carlo wave function simulations of the quantum state evolution  (without normalization) \cite{dalibard_wave-function_1992}
\begin{equation}
\begin{split}
    d|\tilde{\psi}\rangle=&
-\left(
iH+\frac{1}{2}\sum_{n=1,2} \kappa c_n^\dagger c_n
\right)
dt |\tilde{\psi}\rangle\\
&+\sum_{n=1,2}\left(
\sqrt{\kappa dt} {c_n}-I
\right)dN_n |\tilde{\psi}\rangle,
\end{split}
\end{equation}
where $dN_{n}=0,1$ denotes the counting signal, which is a stochastic Poisson increment. The CFI can then be obtained as the mean value over many trajectories, $F_{\theta}=\sum_{j}p_{j}(\partial_{\theta}\ln p_{j})^{2}$, where $p_{j}=\langle \tilde{\psi}|\tilde{\psi}\rangle$ is the probability for each specific trajectory, see  Ref.~\cite{gammelmark_bayesian_2013} for details of the calculation and for the implementation with homodyne detection. An important difference between the CFI and the QFI is that the CFI depends on the choice of measurements as they are accompanied by different measurement back action. In particular, the CFI will differ between direct sensing of the separate signals and sensing of the fields mixed by the beam splitter.

\begin{figure}
    \centering
    \includegraphics[width=1.\linewidth]{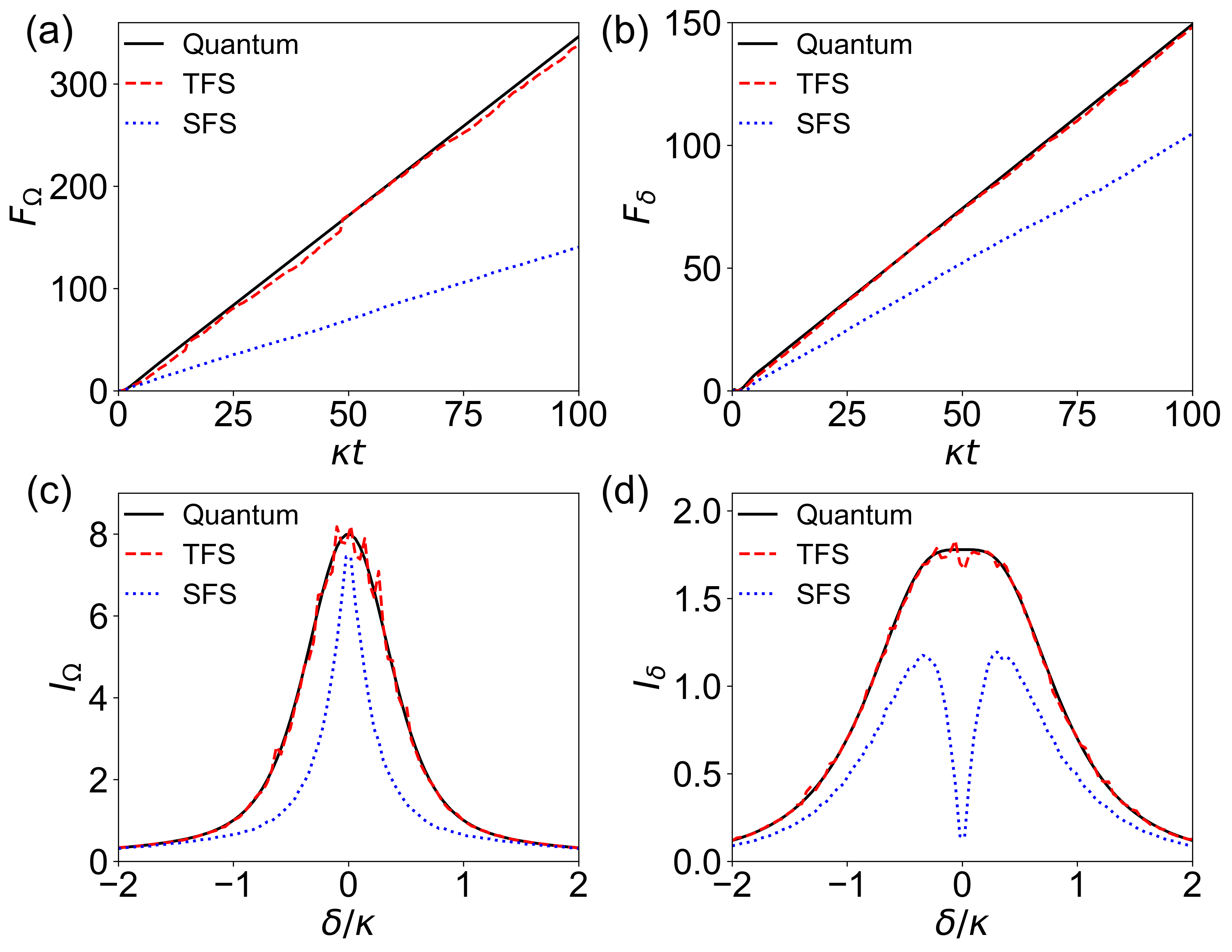}
    \caption{Quantum-limited measurement in two-level atoms. (a),(b) The quantum and two CFI for parameters $\Omega$ (a) and $\delta$ (b). The QFI is denoted by black solid lines (Quantum), and the two CFIs are denoted by blue dotted lines (SFS) and red dashed lines (TFS). The detuning is $\delta/\kappa=0.5$. (c),(d) The corresponding slopes of QFI and two CFIs $I_{\Omega(\delta)}=\lim_{t\to\infty}F_{\Omega(\delta)}/t$. The estimated parameters $\Omega$ and $\delta$ are shared parameters for both systems, i.e., we assume $\delta_{1}=-\delta_{2}\equiv \delta$ and $\Omega_{1}=\Omega_{2}\equiv\Omega$. The dissipation rates are $\kappa=1$, and the Rabi frequency is $\Omega/\kappa=1$. All the CFI is obtained with $10^{4}$ trajectories.}
    \label{fig:2}
\end{figure}

{\it Application to two-level atoms}---The first model we consider is two-level atoms, the Hamiltonians of which are ($i=1,2$)
\begin{equation}\label{eq:atomh}
    H_{i}=-\delta_{i}\sigma_{i}^{+}\sigma_{i}^{-}+\frac{\Omega_{i}}{2}(\sigma_{i}^{+}+\sigma_{i}^{-}),
\end{equation}
where $\sigma_{i}^{\pm}$ are Pauli matrices and $\delta_{i}$ and $\Omega_{i}$ are the detuning and Rabi frequency of the atoms. The corresponding jump operators are $c_{i}=\sigma_{i}^{-}$.

Here we consider the case for real parameters, $\delta_{1}=-\delta_{2}\equiv \delta$ and $\Omega_{1}=\Omega_{2}\equiv\Omega$.
In Figs. \ref{fig:2}(a) and \ref{fig:2}(b), we show excellent agreement between the QFI and the CFI of the TFS protocol for the estimation of $\Omega$ and $\delta$, respectively. For comparison, the lower curves show the total CFI for the SFS. {All the CFI is obtained with $10^{4}$ trajectories.}

In the long-time limit, the CFI and QFI are both proportional to time. We further compare their slopes with respect to time [Figs. \ref{fig:2}(c) and \ref{fig:2}(d)].
The slope of the QFI is directly obtained by differentiating the eigenvalues of the generalized master equation Eq. \eqref{eq:gm} \cite{gammelmark_fisher_2014}, while the slope of the CFI is obtained by a linear fit in a fixed time range $20<\kappa t<100$, causing the fluctuations of CFI in the numerical results (colored lines). The slope of the CFI for the TFS matches the slope of the QFI for all detunings, while for the SFS we observe the known off-resonance suppression of the sensing of the Rabi frequency, and an even more marked suppression for the sensing of the detuning near resonance $\delta=0$.

{\it Application to cavity QED systems}---We further illustrate the performance of the TFS protocol in optical and hybrid systems, the Hamiltonians of which are ($i=1,2$)
\begin{equation}\label{eq:op}
    H_{i}^{\prime}=-\Delta_{i}a_{i}^{\dagger}a_{i}+\varepsilon_{i}a_{i}^{\dagger}+\varepsilon_{i}^{*}a_{i},
\end{equation}
\begin{equation}
    H_{i}^{\prime\prime}=H_{i}^{\prime}-\delta_{i}\sigma_{i}^{+}\sigma_{i}^{-}+\lambda_{i}(a_{i}^{\dagger}+a_{i})(\sigma_{i}^{+}+\sigma_{i}^{-}),
\end{equation}
respectively, where $a_{i}$ are the annihilation operators, $\Delta_{i}$, $\varepsilon_{i}$ are detunings, driving strength, $\lambda_{i}$ is the photon-atom coupling strength, and $a_{i}$ are the annihilation operators. We only consider the optical dissipation and neglect the spontaneous emission of the spin in the hybrid model, so the jump operators are both given by $c_{i}^{\prime}=c_{i}^{\prime\prime}=a_{i}$.

\begin{figure}
    \centering
    \includegraphics[width=1.\linewidth]{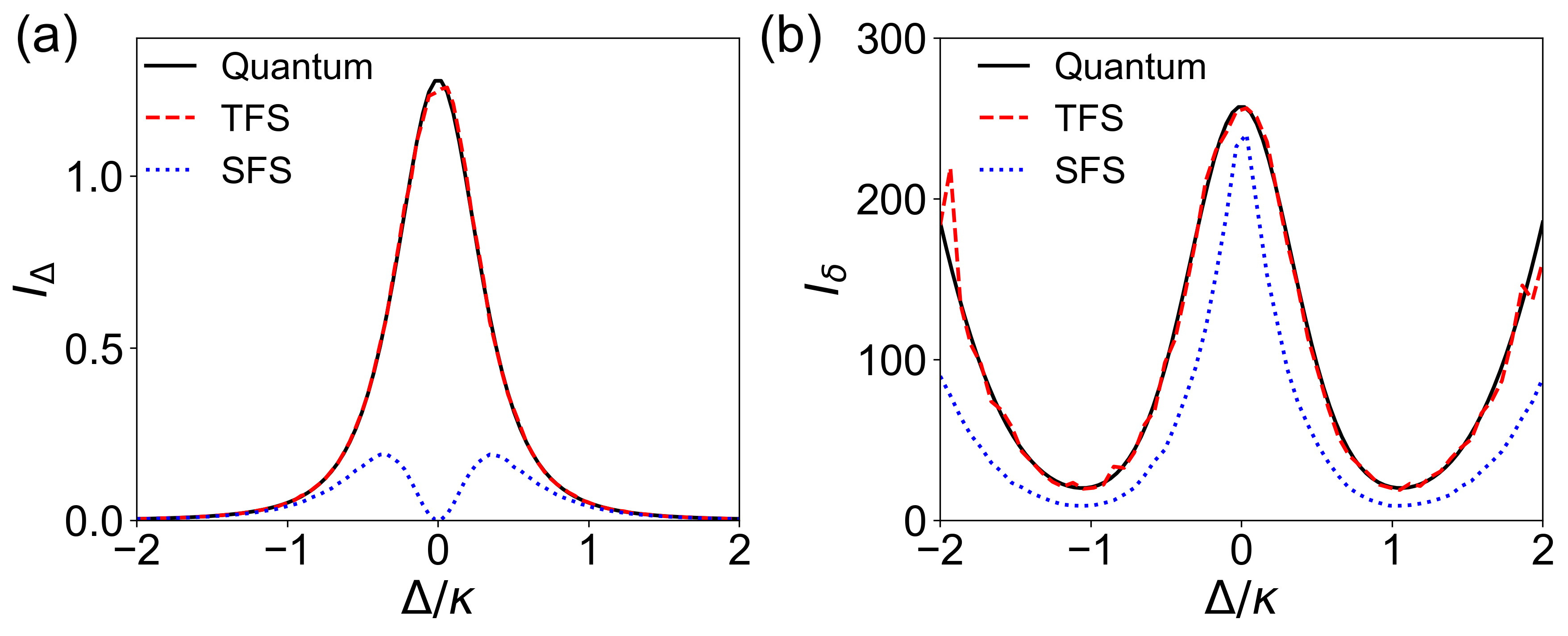}
    \caption{The slopes of the QFI and two CFI when estimating the optical detuning $\Delta$ in optical models (a) and estimating the atomic detuning $\delta$ in hybrid spin-optical models (b).
    The parameters are $\delta/\kappa=1$, $\lambda/\kappa=0.1$ in (b), and $\varepsilon/\kappa=0.1$, $\kappa=1$ in both (a) and (b). The truncated dimension of the Hilbert space is $N_{\mathrm{trun}}=4$ and the number of trajectories is $2\times10^{4}$.}
    \label{fig:3}
\end{figure}

The optical model Eq. \eqref{eq:op} is an optical analog of the two-level emitter, but has an infinite Hilbert space dimension. Similarly, we assume $\Delta_{1}=-\Delta_{2}\equiv\Delta$ and $\varepsilon_{1}=\varepsilon_{2}^{*}\equiv \varepsilon$.
The comparison of the CFI and QFI are shown in Fig. \ref{fig:3}(a). The TFS is far superior to the SFS and saturates the QFI.

{For the hybrid model, we also assume real parameters that satisfy $\Delta_{1}=-\Delta_{2}\equiv\Delta$, $\delta_{1}=-\delta_{2}\equiv\delta$, $\varepsilon_{1}=\varepsilon_{2}^{*}\equiv \varepsilon$, and $\lambda_{1}=\lambda_{2}\equiv \lambda$.
Figure \ref{fig:3}(b) shows the slopes of the CFI and QFI as a function of the optical detuning. The results also verify that the TFS is optimal and can saturate the QFI. Here we consider the driven Rabi model \cite{gietka_unique_2023} because the stationary emission of the Rabi model without driving is the vacuum field \cite{ciuti_input-output_2006}.

{\it Joint sensing of two parameters}---We now show that the TFS can also saturate the QFI for the joint sensing of two parameters. The saturation can be directly proved by the following principle: for a fixed measurement, optimality for two individual parameters implies optimality for their simultaneous estimation. A detailed proof of this principle can be found in the End Matter. In short, since the difference matrix between the quantum and classical Fisher information matrix \cite{liu_quantum_2019,goldberg_intrinsic_2021,wang_achieving_2024} is positive semi-definite, the saturation of all diagonal elements (single-parameter estimations) guarantees the saturation of off-diagonal elements (two-parameter estimation).

An important requirement here is that the saturation of single-parameter estimations must use the same measurement method. The TFS satisfies this requirement as it can use either photon counting or homodyne detection to estimate different system parameters optimally. Consequently, it is also optimal for two-parameter estimation. There may be a formal equivalence between the achievements of TFS for joint sensing of two parameters and the similar achievement in pure-state metrology using antiunitary symmetries \cite{razavian_quantumness_2020,miyazaki_imaginarity-free_2022,wang_achieving_2024}.

{\it Comparison with existing schemes}---
It is interesting that two almost opposite strategies are vigorously pursued in quantum sensing: the negative mass method \cite{polzik_trajectories_2015,tsang_evading_2012,jensen_gaussian_2022,novikov_hybrid_2025} aims to avoid measurement back action as it increases the uncertainty in complementary variables which adds noise to later measurements, while the power of photon counting of resonance fluorescence is precisely due to the quantum jump measurement back action, which quenches the dynamics and makes the subsequent signal more sensitive to unknown parameters. The TFS sensor operates with both elements, as it quenches and entangles the emitters by the measurement back action, while the requirement of a purely imaginary Hamiltonian is obeyed by enforcing opposite and correlated motions in the two systems, as in the negative mass set-up.
The negative mass method is more related to the probe master equation \cite{jacobs_straightforward_2006}, where the continuous measurement is assumed to be performed on the system observables, and thus the dissipative operators are Hermitian dephasing operators instead of non-Hermitian jump operators. Our TFS still applies and can saturate the QFI in this case. Comparison of the performance of the SFS and TFS schemes for this case will be the subject of future investigation.

Our TFS also differs from quantum decoder approaches \cite{yang_efficient_2023} not only in the parallel and cascade structures, but also in the basic mechanism. In the quantum decoder approaches, the auxiliary system mainly acts as a reference or decoder and does not itself encode the unknown parameter. In contrast, the anti-frequency copy in the TFS is an active sensing copy, in which the same parameter is encoded parallel to the sensor. Moreover, in a cascaded decoder architecture, parameter-dependent information can also accumulate in the decoder, leading to non-ergodic dynamics and a growing information component unavailable to the field sensor.

{\it Conclusion}---In conclusion, we have proposed a twin-field sensor (TFS) that enables time-independent continuous measurement to achieve quantum-limited sensitivity. By introducing a suitable copy of the system, we demonstrated that the joint Hamiltonian (evolution operator) can be transformed into a pure-imaginary (pure-real) form. This specific symmetry can fix the optimal measurement basis by merely applying an interference measurement after a 50:50 beam splitter to saturate the quantum limit.
Our numerical simulations across diverse platforms—including driven-dissipative two-level atoms, optical systems, and the Rabi model—confirm that the TFS outperforms conventional SFS, particularly in estimating detuning parameters.
Since this framework applies to all parameters encoded in the Hamiltonian, it is straightforward to verify that it can also saturate the quantum limit for multi-parameter estimation.

These results open several future directions. The TFS method may be extended to noisy sensing scenarios, where loss and environmental noise complicate the construction of optimal measurements \cite{yang_quantum_2026}. It may also provide a useful route for extracting quantum-enhanced information from interacting many-body systems \cite{montenegro_review_2025,somaweera_rydberg_2025,li_exact_2025,zhang_microwave_2026} and designing optimal sensors for distributed quantum metrology \cite{zhuang_distributed_2018,guo_distributed_2020,oh_optimal_2020,zhang_distributed_2021,chen_quantum_2022}.

{\it Acknowledgments}---We thank Weilun Jiang for useful discussions.

%

\newpage
\onecolumngrid
\vspace{30pt}

\begin{center}
\textbf{\large End Matter}
\end{center}    
\twocolumngrid


{\it Unitary transformation of collective operators}---The desired unitary transformation of bosonic collective operator
$a_{\varphi}=(a_{1}+e^{i\varphi}a_{2})/\sqrt{2}$ is given by
\begin{equation}
    U=\exp\left( -i\frac{\pi}{2}a_{\varphi}^{\dagger}a_{\varphi}\right),
\end{equation}
which satisfies
\begin{equation}
    U a_{\varphi} U^{\dagger}=ia_{\varphi},U a_{\varphi+\pi} U^{\dagger}=a_{\varphi+\pi}.
\end{equation}
The independent gauge freedom is easy to understand as $[a_{\varphi}, a_{\varphi+\pi}^{\dagger}]=0$.

Interestingly, the independent gauge freedom still exists for two-level atoms even if the two collective operators do not commute with each other. For simplicity, we consider collective lowering operators $\sigma_{\pm}^{-}=(\sigma_{1}^{-}\pm\sigma_{2}^{-})/\sqrt{2}$, and they satisfy $[\sigma_{+}^{+},\sigma_{-}^{-}]=(\sigma_{1}^{z}-\sigma_{2}^{z})/2$.

The desired unitary transformation has a simple form in the collective basis, i.e., the spin triplet and spin singlet. Assuming the four bases are $|ee\rangle$, $(|ge\rangle+|eg\rangle)/\sqrt{2}$, $(|ge\rangle-|eg\rangle)/\sqrt{2}$, and $|gg\rangle$, the unitary transformation can be written as
\begin{equation}
    U_{-}=\begin{pmatrix}
        1 & 0 & 0 & 0\\
        0 & 1 & 0 & 0\\
        0 & 0 & i & 0\\
        0 & 0 & 0 & 1
    \end{pmatrix},
\end{equation}
and it corresponds to a phase rotation of the asymmetric collective basis state, $(|ge\rangle-|eg\rangle)/\sqrt{2}\to i(|ge\rangle-|eg\rangle)/\sqrt{2}$.
In the same basis, the collective operator can be written as
\begin{equation}
    \sigma_{+}=\begin{pmatrix}
        0 & 0 & 0 & 0\\
        1 & 0 & 0 & 0\\
        0 & 0 & 0 & 0\\
        0 & 1 & 0 & 0
    \end{pmatrix},
    \sigma_{-}=\begin{pmatrix}
        0 & 0 & 0 & 0\\
        0 & 0 & 0 & 0\\
        1 & 0 & 0 & 0\\
        0 & 0 & -1 & 0
    \end{pmatrix}.
\end{equation}
We then have
\begin{equation}
    U_{-}\sigma_{-}^{-}U_{-}^{\dagger}=i\tilde{\sigma}_{-},U_{-}\sigma_{+}^{-}U_{-}^{\dagger}=\sigma_{+},
\end{equation}
where we introduce new collective operators in the rotated basis as
\begin{equation}
    \tilde{\sigma}_{-}=\begin{pmatrix}
        0 & 0 & 0 & 0\\
        0 & 0 & 0 & 0\\
        1 & 0 & 0 & 0\\
        0 & 0 & 1 & 0
    \end{pmatrix}.
\end{equation}

The above discussion can also be expressed using the exchange operators. Defining $P_{\mathrm{sym}}$ and $P_{\mathrm{asym}}$ as the projection operators on the symmetric and asymmetric Hilbert spaces, the unitary transformation can be written as
\begin{equation}
    U_{-}=P_{\mathrm{sym}}+iP_{\mathrm{asym}}.
\end{equation}
It represents the general expression of the unitary transformation that also applies to collective spin operators.

{\it Two-parameter estimation}---We now show that, for a fixed measurement, optimality for two individual parameters implies optimality for their simultaneous estimation. More
precisely, consider a two-parameter model
$\boldsymbol{\theta}=(\theta_1,\theta_2)$ and a fixed measurement
$\mathcal M$, which gives the CFI matrix
$F^C[\mathcal M]$. Let $F^Q$ denote the corresponding quantum Fisher
information matrix. If, at the same parameter point, the same measurement
saturates the two diagonal entries,
\begin{equation}
F^C_{11}=F^Q_{11},\qquad F^C_{22}=F^Q_{22},
\end{equation}
then it also saturates the off-diagonal entry,
\begin{equation}
F^C_{12}=F^Q_{12}.
\end{equation}
Therefore,
\begin{equation}
F^C[\mathcal M]=F^Q,
\end{equation}
for the full two-parameter Fisher information matrix.

To prove this statement, we use the standard multi-parameter
Braunstein--Caves inequality \cite{braunstein_statistical_1994},
\begin{equation}
F^C[\mathcal M]\preceq F^Q,
\end{equation}
or equivalently,
\begin{equation}
D\equiv F^Q-F^C[\mathcal M]\succeq0 .
\end{equation}
For completeness, this matrix inequality can be obtained by considering
an arbitrary one-dimensional sub-model
\begin{equation}
\theta_v=\mathbf v^T\boldsymbol{\theta},
\end{equation}
where $\mathbf v\in\mathbb R^2$. Along this direction, the CFI and
QFI are
\begin{equation}
F^C_{\theta_v}=\mathbf v^T F^C \mathbf v,
\qquad
F^Q_{\theta_v}=\mathbf v^T F^Q \mathbf v .
\end{equation}
The single-parameter Braunstein--Caves inequality gives
\begin{equation}
F^C_{\theta_v}\leq F^Q_{\theta_v},
\end{equation}
for any real vector \(\mathbf v\). Hence,
\begin{equation}
\mathbf v^T(F^Q-F^C)\mathbf v\geq0,
\end{equation}
for all $\mathbf v$, which proves that $D=F^Q-F^C$ is positive
semi-definite.

Consequently, if the same measurement saturates the two single-parameter bounds,
i.e.,
\begin{equation}
D_{11}=F^Q_{11}-F^C_{11}=0,\qquad
D_{22}=F^Q_{22}-F^C_{22}=0.
\end{equation}
Since \(D\succeq0\), its off-diagonal element must satisfy
\begin{equation}
|D_{12}|^2\leq D_{11}D_{22}=0 .
\end{equation}
Therefore,
\begin{equation}
D_{12}=0,
\end{equation}
which implies
\begin{equation}
F^C_{12}=F^Q_{12}.
\end{equation}
Thus, the same measurement saturates the full two-parameter quantum Fisher information matrix.


\begin{figure}[t]
    \centering
    \includegraphics[width=1.\linewidth]{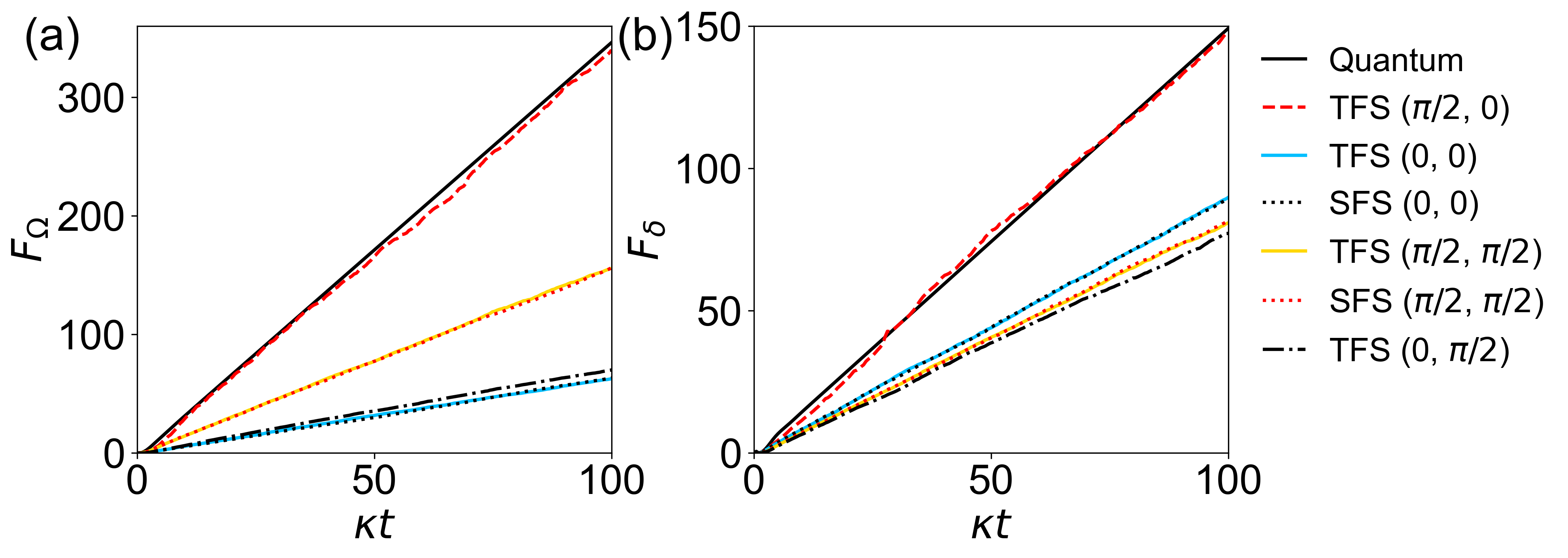}
    \caption{The QFI and CFI of different homodyne detections for parameters $\Omega$ (a) and $\delta$ (b). The order in the labels corresponds to the position of the lines, from top to bottom. The phases in the labels denote the homodyne phases in two measured channels. For the SFS (TFS), the two phases correspond to channels $c_{1}$ $(c_{+})$ and $c_{2}$ $(c_{-})$, respectively. The detuning is $\delta/\kappa=0.5$, the dissipation rates are $\kappa=1$, and the Rabi frequency is $\Omega/\kappa=1$.}
    \label{fig:s1}
\end{figure}

{\it Homodyne detection}---Here, we further verify that the TFS also works for homodyne detection. We consider two-level atoms as an example, and we compute the CFI through homodyne detection for both SFS and TFS. As shown in Fig. \ref{fig:s1}, we compute two situations for SFS and four situations for TFS, and only the homodyne detection in TFS with homodyne phases $(\pi/2, 0)$ can saturate the QFI. It corresponds to the joint measurements of field quadratures $p_1+p_2$ and $x_1-x_2$, the same as in the negative-mass methods \cite{polzik_trajectories_2015}. In contrast, when measuring $p_1-p_2$ and $x_1+x_2$ [TFS $(0, \pi/2)$], the performance of the TFS is even worse than that of the optimal SFS. In the case of the same homodyne phases $(0, 0)$ or $(\pi/2, \pi/2)$, the CFI of the TFS is the same as that of the SFS, as the measurement of $x_{1}\pm x_{2}$ is equivalent to the separate measurement of $x_{1}$ and $x_{2}$.

\end{document}